\documentclass[10pt,conference]{IEEEtran}
\usepackage{epsfig,setspace,amsmath,epsf,amssymb,bm,cite,graphicx, epstopdf,algorithm,algpseudocode,float,mathtools,authblk,tikz,physics,amsthm}
\usetikzlibrary{arrows.meta,positioning,decorations.pathmorphing,calc,shapes.geometric}
\usepackage[short,c2]{optidef}

\usepackage{empheq}
\usepackage{cases}

\usepackage[table,xcdraw]{xcolor}
\usepackage{booktabs}
\usepackage{subfigure}
\usepackage{bbm}

\IEEEoverridecommandlockouts
\allowdisplaybreaks
\begin{document}

\title{Squeezing the Most Out of Preemption \\ for AoI Minimization: Single-source Case}
\author[1]{Nail Akar}
\author[2]{Mohammad Moltafet}
\author[3]{Sennur Ulukus}
\author[4]{Marian Codreanu}
\author[5]{Roy D. Yates}

\affil[1]{Bilkent University, Ankara, T\"{u}rkiye}
\affil[2]{Tennessee Tech University, TN, USA}
\affil[3]{University of Maryland, College Park, USA}
\affil[4]{Link\"oping University, Link\"oping, Sweden}
\affil[5]{Rutgers University, New Brunswick, NJ, USA}

\maketitle
\begin{abstract}
In this work, we study a single-source single-server continuous-time status update system where the updates arrive according to a Poisson process and update service times are generally distributed. 
In our proposed setting, a preemption policy refers to one where a new update preempts the ongoing one with a probability depending on the age of the update in service. We first propose an analytical method to derive the average age of information (AoI) and average peak AoI (PAoI) for any such preemption policy.  This analysis is then utilized to tune two particular preemption policies: (i) probabilistic preemption (PP), in which preemption takes place according to a fixed probability regardless of the update age, (ii) threshold-based preemption (TP), for which preemption is incurred when the update age exceeds a certain threshold, both using one-dimensional line search. The effectiveness of policy tuning for the PP and TP policies is validated using lognormal-distributed update service times. 
\end{abstract}

\section{Introduction}
\label{section1}
Information freshness has become a key concept for successful operation of networked control and monitoring systems. In this paper, we consider a monitoring system comprising an information source and a remote monitor, separated from each other by a network. At times, the source samples its associated random process and sends the updates through the network, which introduces a random delay that is modeled by a server with generally distributed service times.
Among several metrics, age of information (AoI) has recently gained significant attention, as an information freshness metric, thanks to its obliviousness to the dynamics of the source process \cite{kaul_etal_infocom12,kosta_etal_survey,yates_survey}.
The AoI process held at the monitor keeps track of the time elapsed since the generation of the last received update. Consequently, the AoI process is a cyclic process which increases in time with a unit slope within a cycle, but is subject to an abrupt downward jump at update reception instances, upon which a new cycle begins. In our proposed setting, the AoI process is an asymptotically stationary and ergodic continuous-time, continuous-valued stochastic process, and our interest in this paper lies in finding its time average. 
On the other hand, the discrete-time, continuous-valued peak AoI (PAoI) process is obtained by sampling the AoI process just before update receptions \cite{costa_peak}, and we are interested in finding its time average as well.

Although status update systems with multiple sources and servers have been studied in the literature,
we focus our attention in this paper on a single-source single-server update system where the update arrivals are Poisson, and update service times (or network delays) are independent and identically distributed. No particular distribution is assumed for the update service times which are assumed to have arbitrary distributions. Particularly, we study preemption policies where a new update preempts the ongoing one with a certain probability depending on the age of the update in service which is defined as the elapsed time in service for the ongoing update. 
The situation in which preemption can also be based on the instantaneous AoI process is left for future research.
We propose a method to derive the average AoI and average PAoI for any such preemption policy. The proposed analytical method is then used to tune two preemption policies parameterized by a single parameter, by using  one-dimensional line search. The first policy is the so-called probabilistic preemption (PP) policy, where preemption takes place according to a fixed probability which is oblivious to update age \cite{moltafet_etal_isit25}. The second policy is threshold-based preemption (TP), where the new arriving update preempts the ongoing one once the update age exceeds a given threshold.

Our main contributions are given below.
\begin{itemize}
    \item We propose a method to derive the statistics of the lifetime of updates for generally distributed service times and update age-dependent preemption, using the theory of non-homogeneous absorbing CTMCs, which to the best of our knowledge, is novel. 
    \item The update lifetime statistics are then used to derive the average AoI and PAoI for the status update system.
    \item Finally, we use the proposed analytical model to tune two preemption policies, namely probabilistic preemption and threshold-based preemption, both of which are characterized by a single parameter. Notable reductions in average AoI and PAoI are observed with the use of these two policies, provided they are tuned properly.
\end{itemize}
The paper is organized as follows. 
Related work is briefed in Section~\ref{sec:related}.
Section~\ref{sec:AoI} formally describes
the AoI and PAoI processes for a generic status update system. 
In Section~\ref{sec:system}, we describe the system model. Section~\ref{sec:analytical} presents the analytical method. Numerical results are presented in Section~\ref{sec:numerical}. We conclude in Section~\ref{sec:conclusions}.

\section{Related Work}
\label{sec:related}
The average AoI is first obtained in \cite{kaul_etal_infocom12} for single-source M/M/1, M/D/1, and D/M/1 queues with infinite buffer capacity and first come first serve (FCFS) scheduling, whereas
\cite{costa_etal_TIT16} investigates the distribution of AoI and PAoI for the bufferless M/M/1/1 and the single-buffer M/M/1/2 systems, along with the M/M/1/2$^{\ast}$ model where the packet waiting in the queue is to be replaced by a fresh packet arrival. \cite{inoue_etal_IT19} presents results for the steady-state distributions of AoI and PAoI for a general class of single-source systems.

There has also been substantial interest on preemptive status update systems. We first review the existing work on single-source systems.
The average AoI and PAoI for the preemptive last come first serve (LCFS) M/G/1/1 queue is studied in \cite{najm_nasser_isit16} where the service time is modeled by the gamma distribution.
\cite{akar_etal_tcom20} investigates the distributions of AoI and PAoI in single-source bufferless systems with probabilistic preemption while allowing  
phase type  distributions for both inter-arrival and service times. 
The reference \cite{soysal_ulukus_IT21} derives upper bounds for the mean AoI for
the G/G/1/1 queue as well as its preemptive version while showing that the bounds are close to actual values. The work in \cite{moltafet_etal_isit25} studies the average AoI and PAoI for a preemptive server with generally distributed service times and presents advantages of probabilistic preemption for certain service time distributions. In \cite{li_2026_tamin}, the authors consider a single-source generate-at-will status update system with generally distributed service times. They study a continuous-time joint sampling and preemption problem to minimize the AoI.

Studies in the multi-source setting are now briefed. 
The work of \cite{najm_telatar_infocom18} investigates an M/G/1/1 queue with preemption and computes a closed form expression for the average AoI and PAoI of each source, by using the detour flow graph.
When service time distributions are phase type, the authors of \cite{dogan_akar_tcom21} provide an algorithm to obtain the distributions of AoI and PAoI for a multi-source M/PH/1/1 model under probabilistic preemption.
\cite{moltafet_etal_tcom22} investigates a multi-source M/G/1/1 queuing model considering a self-preemptive packet management policy, and derives the moment generating function (MGF) of AoI and PAoI.

\section{AoI and PAoI }
\label{sec:AoI}
\begin{figure}[tb]
    \centering
    \begin{tikzpicture}[scale=0.25]	
    \draw[<->,black,thick] (9,13) -- (16,13);
    \filldraw (12.5,15) circle (0.01) node[anchor=south, thick] {cycle-$k$};
     \filldraw (12.5,13) circle (0.01) node[anchor=south, thick] {$V_k$};
\draw[<->,black,thick] (16,13) -- (24.4,13);
    \filldraw (21,15) circle (0.01) node[anchor=south, thick] {cycle-$(k+1)$};
    \filldraw (21,13) circle (0.01) node[anchor=south, thick] {$V_{k+1}$};
    \draw[thick,->] (1,0) -- (30,0) node[anchor=north] {$t$};
    \draw[thick,->] (1,0) -- (1,14) node[anchor=south] {$\Delta(t)$};
    \draw[ultra thick,blue] (4.5,4.5) -- (9,9);
    \draw[dashed,very thin] (4.5,4.5) -- (9,9);
    \filldraw[blue] (2.5,2.5) circle (3pt);
    \filldraw[blue] (3.5,3.5) circle (3pt) ;
    \filldraw[blue] (3,3) circle (3pt); 
    \draw (1,5) node[anchor=east] {$U_k$};
     \draw[dotted,gray] (1,12) -- (30,12);
    \draw[dotted,gray] (1,9) -- (30,9);
    \draw[dotted,gray] (1,5) -- (30,5);
    \draw[dotted,gray] (1,2) -- (30,2);
    \draw[dotted,gray] (1,10.5) -- (30,10.5);
\draw[dotted,gray] (1,3.5) -- (30,3.5);
    
    \draw[dotted,gray,thick] (9,12.5) -- (9,0)  node[anchor=north, thick, black] {$R_{k}$};
    \draw[ultra thick,blue] (9,9) -- (9,5.2);
    \draw[ultra thick,blue] (9.1,5.1) -- (16,12);
    \draw[dotted,gray,thick] (16,12) -- (4,0)  node[anchor=north, thick, black] {$G_{k}$};
    \node at (3.75,0.50) (nodeA) {};
    \node at (8.75,5.5) (nodeB) {};
    \node at (15.75,12.5) (nodeC) {};
    
    \draw (1,12) node[anchor=east] {$\Phi_k$};
    \draw (1,9) node[anchor=east] {$\Phi_{k-1}$};
     \draw (1,10.5) node[anchor=east] {$\Phi_{k+1}$};
    \draw (1,2) node[anchor=east] {$U_{k+1}$};
     \draw (1,3.5) node[anchor=east] {$U_{k+2}$};

    \draw[dotted,gray,thick] (16,12.5) -- (16,0)  node[anchor=north, thick, black] {$\quad R_{k+1}$};
    \draw[ultra thick,blue] (16,12) -- (16,2.2);
    \draw[ultra thick,blue] (16.1,2.1) -- (24.5,10.5);
    \draw[dotted,gray,thick] (20,6) -- (14,0)  node[anchor=north, thick, black] {${G_{k+1}} \quad$};

    \draw[ultra thick,blue] (24.5,10.5) -- (24.5,3.5);
    \draw[ultra thick,blue] (24.6,3.6) -- (27.5,6.5);
     \filldraw[blue] (28.5,7.5) circle (3pt);
    \filldraw[blue] (29,8) circle (3pt) ;
    \filldraw[blue] (29.5,8.5) circle (3pt);

    \draw[dotted,gray,thick] (24.5,3.5) -- (24.5,0)  node[anchor=north, thick, black] {$\quad R_{k+2}$};
    \draw[dotted,gray,thick] (24.5,3.5) -- (21,0)  node[anchor=north, thick, black] {${G_{k+2}} \quad$};

    \draw[gray] (9,5) circle (6pt);
    \draw[gray] (16,2) circle (6pt);
     \draw[gray] (24.5,3.5) circle (6pt);
    \end{tikzpicture}
    \caption{Sample path of the AoI process $\Delta(t)$. Only generation and reception instances of successful packets are shown.}
    \label{fig:samplepath}
\end{figure}
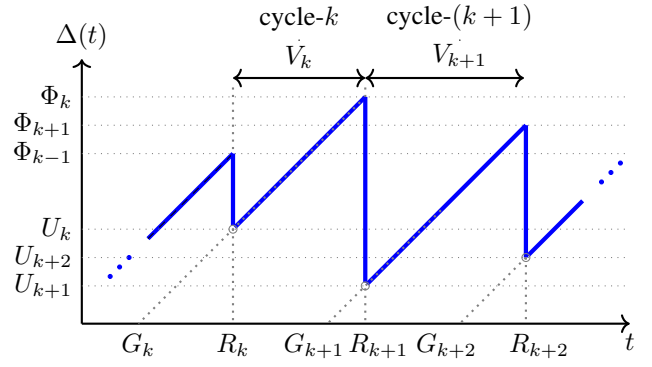

We describe the AoI and PAoI processes for a generic status update system. Let $G_k$ and $R_k$ denote the instances at which the $k^{\text{th}}$ update is generated by the source and received by the monitor, respectively. Let $U_k$ denote the system time of the $k^{\text{th}}$ successful update, i.e., $U_k=R_k - G_k$. Fig.~\ref{fig:samplepath} depicts a sample path of the AoI process $\Delta(t)$ (thick blue solid curve). During cycle-$k$,  $\Delta(t)$ increases with unit slope from the value $U_k$ at time $R_{k}$ until the value $\Phi_k$ at time $R_{k+1}$, when it abruptly drops to $U_{k+1}$. On the other hand, the PAoI process $\Phi_k, k \geq 0$ is obtained by sampling the AoI process $\Delta(t)$ at the embedded time points just before receptions.
Note that the AoI and PAoI processes are asymptotically stationary due to the random nature of the service times.
Consequently, we let $\Delta$ and $\Phi$ denote the steady-state random variables for the random process $\Delta(t)$ and $\Phi_k$, respectively, with cdf 
$F_{\Delta}(x) = \lim_{t \rightarrow \infty} \mathbb{P} ( \Delta(t) \leq x)$ and $F_{\Phi}(x) = \lim_{k \rightarrow \infty} \mathbb{P} ( \Phi_k \leq x)$ for $x \geq 0$. Hence,
\begin{align}
    \mathbb{E}[\Delta] & = \lim_{T \rightarrow \infty} \frac{1}{T} \int_{t=0}^T \Delta(t) \dd{t}, \label{averageAoI} \\
    \mathbb{E}[\Phi] & = \lim_{K \rightarrow \infty} \frac{1}{K} \sum_{k=1}^K \Phi_k, \label{averagePAoI}
\end{align}

\section{System Model}
\label{sec:system}
\begin{figure*}
\centering
 \resizebox{0.7\linewidth}{!}{
\begin{tikzpicture}[
    >=Stealth,
    thick,
    font=\small,
    block/.style = {draw, rounded corners=2pt, minimum width=2.6cm,
                     minimum height=1.3cm, align=center, fill=blue!4},
    src/.style   = {draw, circle, minimum size=1.1cm, align=center, fill=orange!12},
    dst/.style   = {draw, circle, minimum size=1.1cm, align=center, fill=green!10},
    dec/.style   = {draw, align=center, font=\scriptsize}, 
]

\node[src] (src) {Source};

\node[block, right=2.3cm of src] (srv)
      {Server\\[1pt] \footnotesize hazard rate $q(x)$};

\node[dst, right=2.3cm of srv] (dst) {Monitor};

\draw[->] (src) -- node[above, font=\scriptsize]
      {Poisson$(\lambda)$} (srv);

\draw[->] (srv) -- node[above, font=\scriptsize]{completed update} (dst);

\node[dec, above=1.10cm of srv, xshift=0cm] (rule)
      {new arrival preempts update\\in service (age $x$) w.p. $p(x)$};

\draw[->, dashed] (rule.south) -- (srv.north);
\draw[-{Stealth[scale=0.9]}, dashed]
      (srv.north) to[bend right=25] node[left, font=\scriptsize, xshift=-2pt]
      {} (rule.south west);

\node[dec, below=0.55cm of srv, gray, text=black] (disc)
      { arriving update is discarded w.p. $1-p(x)$ };
\draw[->, gray] (srv.south) -- (disc.north);

\end{tikzpicture}
}
\caption{Status update system involving a single source, a server with general service time characterized with hazard rate $q(x)$ and buffer management policy characterized with preemption probability $p(x)$, and a remote monitor. }
\label{fig:systemmodel}
\end{figure*}
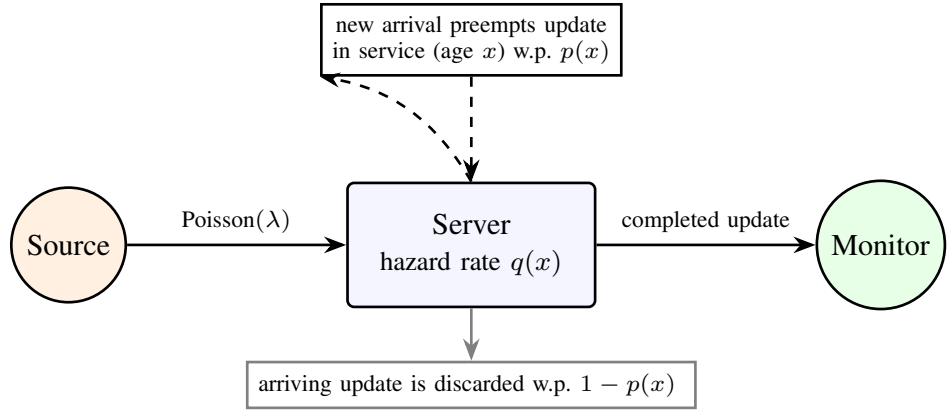

We consider a status update system involving a single source and a monitor in Fig.~\ref{fig:systemmodel}. The source sends update messages (or updates) towards the monitor by receiving service from a single server. The updates are generated according to a Poisson process with rate $\lambda$. The service time for an update, denoted by $S$, is generally distributed with pdf $f(x)$, cdf $F(x)$, and the hazard rate
$q(x)$ which is defined as,
\begin{align}
    q(x) & = \lim_{h \rightarrow 0} \frac{1}{h} \Pr (x < S \leq x + h \ | \ S > x) = \frac{f(x)}{1-F(x)},  
\end{align}
which is also known as the instantaneous conditional completion rate when the elapsed time of the service time (or update age) has the value $x$. The cumulative hazard function, denoted by $Q(x)$ is given by,
\begin{align}
    Q(x) & = \int _0^x q(t) \dd{t} = - \ln (1-F(x)).
\end{align}
For the hazard rates and cumulative hazard functions of well-known distributions and also their definitions, we refer the reader to \cite{book_oconnor}.
When a new update message arrives, it immediately starts to receive service if the server is idle. When the server is busy serving an update when the new update arrives, we assume that the new update preempts the ongoing one with probability $p(x)$ when the instantaneous age of the update is $x$. 
This general preemption policy contains some well-known preemption policies as its sub-cases.
When $p(x)=1$ for all $x$, then we have a fully preemptive (FP) policy \cite{dogan_akar_tcom21}.
On the other hand, the case of $p(x)=0$ for all $x$ is called the non-preemptive (NP) policy \cite{dogan_akar_tcom21}.
The probabilistically preemptive (PP) policy of \cite{moltafet_etal_isit25} uses a fixed probability $\alpha$ for preemption, i.e., $p(x)=\alpha$ for all $x$.
In this paper, we propose to use a threshold-based preemptive (TP) policy for which $p(x)=1$ for $x > \tau$ for a given threshold $\tau$, and is otherwise zero. The motivation behind the TP policy is that when the update age is relatively large, it may be advantageous to preempt it with a fresh packet since otherwise the next AoI cycle would begin from a very large value, despite the fact that the current packet has been served for some time. However, the optimum value of $\tau$ needs to be determined that will yield the minimum average AoI or PAoI. 

Our first goal is to devise a method to derive the average AoI and peak AoI for the update system characterized with the 3-tuple $(\lambda,q(x),p(x))$, which allows one to study a wide range of preemption policies.
Our second goal is to use this general method to derive the optimum PP and TP policies.
Since TP is characterized by a single parameter $\tau$ similar to PP which is characterized by a single probability $\alpha$, line search techniques can be used for both to solve the optimum parameters $\tau$ and $\alpha$, respectively.
\section{Analytical Method}
\label{sec:analytical}
The analytical method proposed for the update system (characterized by the 3-tuple $(\lambda,q(x),p(x))$)
consists of the following two steps. In the first step, we study the lifetime of a single update that receives service. Using this update lifetime statistics, we derive the average AoI and PAoI in the second step.
\subsection{Lifetime of an Update}
Let us consider an update $u$ which receives service.
The lifetime of the update $u$, denoted by $T$, is defined as the duration of time $u$ receives service. This service will be over either because $u$ is preempted with a new update, or the service completes on its own. In order to study $T$, we construct an absorbing Markov chain (AMC)  $Y(x)$ with one transient state $0$, and two absorbing states, namely preemption state $1$ and completion state $2$. 
The AMC $Y(x)$ starts operation with the update $u$ starting to receive service at time $x=0$. $Y(x)$ stays in the transient state $0$ until absorption occurs into the state $1$ when $u$ is preempted with a new update arrival, or absorption occurs into state $2$ when the service of $u$ is complete without being preempted by another update. Therefore, the lifetime $T$ amounts to the absorption time of the AMC $Y(x)$. The non-homogeneous generator for the AMC $Y(x)$, denoted by $G(x)$ is easy to write:
\begin{align}
G(x) & = \begin{pmatrix}
-\left( q(x) + \lambda p(x) \right) & \lambda p(x) & q(x) \\
0 & 0 & 0 \\
0& 0 & 0 
\end{pmatrix}.
\end{align}
The probability that the AMC $Y(x)$ is at state $0$ at time $x$, denoted by $\pi(x)$, is written as,
\begin{align}
     \pi(x) & =  \Pr (T > x) = {\rm e}^{- \int_0^x \left( q(t) + \lambda p(t) \right)  \dd{t}},
\end{align}
which is the unique solution to the first-order differential equation,
\begin{align}
    \dv{x} \pi(x) &= - \left( q(x) + \lambda p(x) \right)\pi(x), \quad x \geq 0.
\end{align}
Let $p$ and $q$ denote the absorption probability into the absorbing states $1$ and $2$, respectively. We can thus write,
\begin{align}
    p & = \Pr(Y(\infty)=1) = \int_0^{\infty} \pi(x) \lambda p(x) \dd{x}, \\
    q & = \Pr(Y(\infty)=2) = \int_0^{\infty} \pi(x) q(x) \dd{x}.   
\end{align}
Let $T_p$ and $T_q$ denote the lifetime of the update $u$ conditioned on absorption into the preemption state $1$ and completion state $2$, respectively. We can write the pdf of $T_p$ as,
\begin{align}
    f_{T_p}(x) & = f_{T |Y_1}(x) = \frac{1}{p} \pi(x) \lambda p(x),
\end{align} from which one can obtain the first two moments of $T_p$, namely $\mathbb{E}[T_p]$ and 
$\mathbb{E}[T_p^2]$. Here, $Y_1$ refers to the event $\{ Y(\infty)=1 \}$.
Similarly, we write the pdf of $T_q$ as,
\begin{align}
    f_{T_q}(x) & = f_{T | Y_2}(x) = \frac{1}{q} \pi(x) q(x).
\end{align} by means of which we can obtain $\mathbb{E}[T_q]$ and 
$\mathbb{E}[T_q^2]$. Similarly, $Y_2$ refers to the event $\{ Y(\infty)=2 \}$.
\subsection{Derivation of Average AoI/PAoI}
Consider the AoI cycle-$k$ given in Fig.~\ref{fig:samplepath} for which $\Delta(t)$ rises up from the value $U_k$ to $\Phi_k = U_k + V_k$ in a time duration of $V_k$. Note that $U_k \sim U$ and $V_k \sim V$ where $U$ and $V$ stand for the random variables corresponding to the system time of received updates, and the duration between two successful service completions, respectively. 
Here, $A  \sim B$ when $A$ and $B$ have the same distribution, and further note that $U$ and $V$ are independent.
The mean AoI $\mathbb{E}[\Delta]$ is then written as the ratio of the area under the AoI curve in one cycle to the duration of the cycle \cite{yates_survey},
\begin{align}
  \mathbb{E}[\Delta]  & = \mathbb{E}[U] + \frac{\mathbb{E}[V^2]}{2 \mathbb{E}[V]}. \label{meanAoIexp}
  \end{align}
On the other hand, the mean PAoI $\mathbb{E}[\Phi]$ is written as
\begin{align}
    \mathbb{E}[\Phi] & = \mathbb{E}[U] + \mathbb{E}[V]. \label{meanPAoIexp}
\end{align}
Let us first note that $U \sim T_q$ since $U$ amounts to the system time of a successful update. Hence, \begin{align}
    \mathbb{E}[U] & = \mathbb{E}[T_q]. \label{meanofU}
    \end{align}
The study of the random variable $V$ is more involved. Actually, $V$ is the sum of three random variables,
\begin{align}
    V &= V_a + V_p + V_q, \label{sumofthree}
\end{align}
where $V_a$ is the time needed until the first update arrival, 
$V_p$ is the duration of all update lifetimes ending up with preemption, and $V_q$ is the time needed for a successful update.
Since $V_a$ is exponentially distributed with parameter $\lambda$, we have
\begin{align}
    \mathbb{E}[V_a] & = \frac{1}{\lambda}, \quad \text{Var}(V_a) = \frac{1}{\lambda^2}.
\end{align}
On the other hand,
$V_p$ is written as the following sum,
\begin{align}
    V_p & = \sum_{i=1}^K T_{p,i},
\end{align}
where $K$ has a delayed geometric distribution with parameter $p$ and $T_{p,i}$s are independent of each other and  $T_{p,i} \sim T_p$.
The probability generating function (pgf) of $K$, denoted by $G_K(z)=\mathbb{E}[z^K]$, can be written as,
\begin{align}
    G_K(z) & = q + p q z + p^2 q z^2 + \cdots, \\
           & = \frac{q}{1-pz}.
\end{align}
Also, let us define the moment generating functions (mgf) $H_{V_p}(s) = \mathbb{E}[{\rm e}^{s V_p}]$ and
$H_{T_p}(s) = \mathbb{E}[{\rm e}^{s T_p}]$.
Using the method of collective marks \cite{kleinrock_book}, we can write,
\begin{align}
    H_{V_p}(s) & = G_K\left(H_{T_p}(s)\right), \\
            & = \frac{q}{1-p \left( \sum_{i=0}^\infty \mathbb{E}[T_p^i] \frac{s^i}{i!} \right)},  \\
               & =  \frac{q}{1-p \left(1 + \mathbb{E}[T_p]s + \mathbb{E}[T_p^2]\ \frac{s^2}{2} + R(s)\right) },\label{mgf}
\end{align}
where the first two derivatives of $R(s)$ vanish at zero.
By differentiating the expression \eqref{mgf} with respect to $s$ and evaluating the expression at $s=0$, we obtain,
\begin{align}
    \mathbb{E}[V_p] & = \eval{\dv{s}  H_{V_p}(s)}_{s=0} = \frac{p \mathbb{E}[T_p]}{1-p}.
\end{align}
By differentiating the expression \eqref{mgf} once more with respect to $s$ and evaluating the expression at $s=0$, we obtain 
\begin{align}
    \mathbb{E}[V_p^2] & = \eval{\dv[2]{s}  H_{V_p}(s)}_{s=0}, \\
                      & = \frac{2 p^2 \mathbb{E}[T_p]^2}{(1-p)^2} + \frac{p \mathbb{E}[T_p^2]}{1-p}, \\
                \text{Var}(V_p) & =   \mathbb{E}[V_p^2] -    \mathbb{E}[V_p]^2.   
\end{align}
Finally, $V_q$ is the time needed for the service of a successful update, i.e., $V_q \sim T_q$. 
Since $V_q \sim T_q$, we can write 
\begin{align}
    \mathbb{E}[V_q] & =  \mathbb{E}[T_q], \quad \text{Var}(V_q) = \mathbb{E}[T_q^2] - \mathbb{E}[T_q]^2.
\end{align}
Finally, we write from \eqref{sumofthree} the following,
\begin{align}
    \mathbb{E}[V] & = \mathbb{E}[V_a] + \mathbb{E}[V_p] + \mathbb{E}[V_q], \label{meanofV}
 \\
 \text{Var}(V) & = \text{Var}(V_a) + \text{Var}(V_p) + \text{Var}(V_q), \label{varianceofthree} \\
 \mathbb{E}[V^2] & =  \text{Var}(V) + \mathbb{E}[V]^2. \label{secondmomentofV}
 \end{align}
 Finally, we obtain the average AoI $\mathbb{E}[\Delta]$ and the average PAoI $\mathbb{E}[\Phi]$ by plugging in the identities \eqref{meanofU}, \eqref{meanofV}, and \eqref{secondmomentofV} in the equations \eqref{meanAoIexp} and \eqref{meanPAoIexp}, respectively.  
 \section{Numerical Results}
 \label{sec:numerical}
\begin{figure*}[tb]
    \centering
    \includegraphics[width=0.7\linewidth]{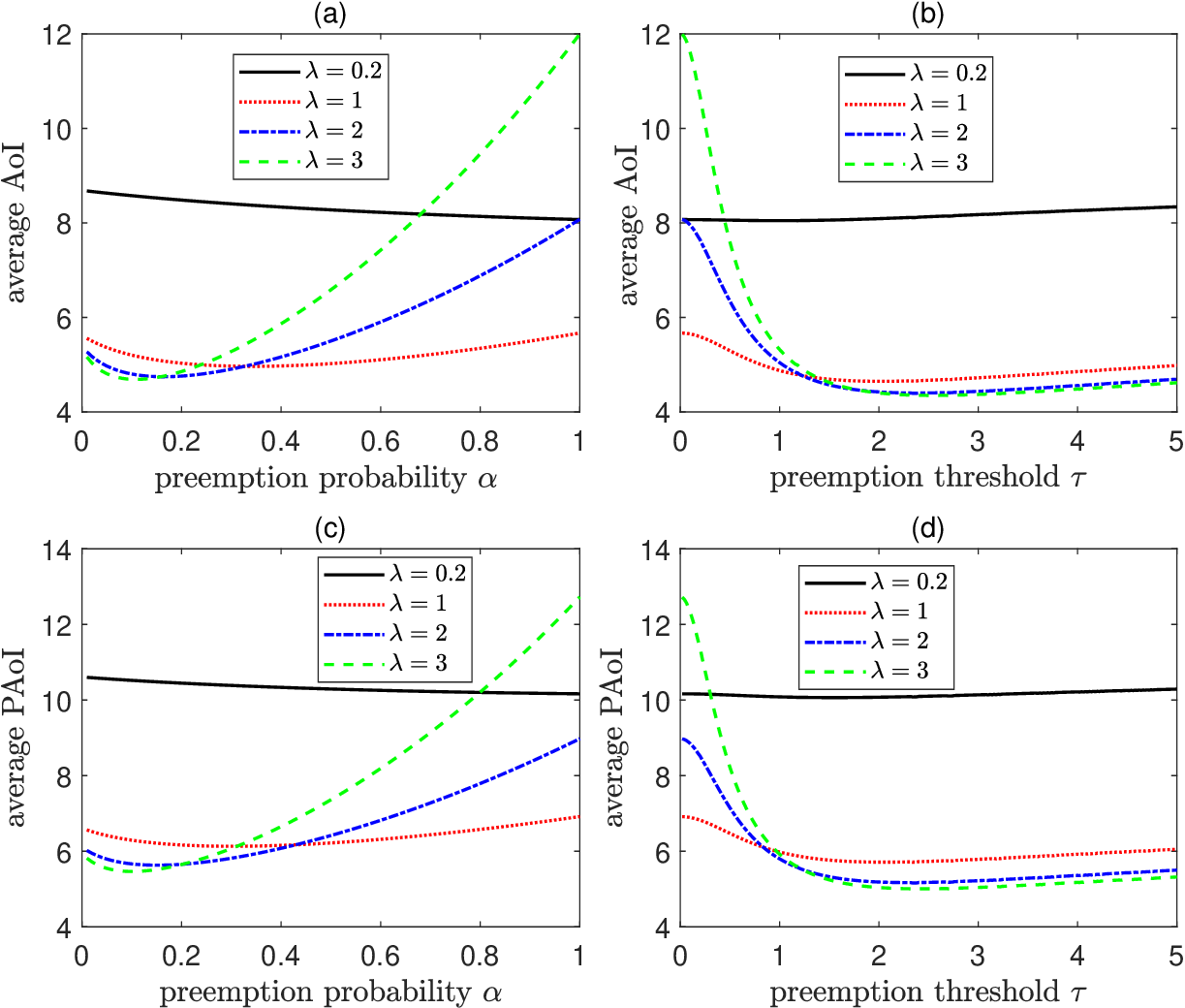}
    \caption{(a) Average AoI vs $\alpha$ (PP) (b) Average AoI vs $\tau$ (TP) 
    (c) Average PAoI vs $\alpha$ (PP) (b) Average PAoI vs $\tau$ (TP) }
    \label{fig:PaperFig1}
\end{figure*}

\begin{figure*}[tb]
    \centering
    \includegraphics[width=0.7\linewidth]{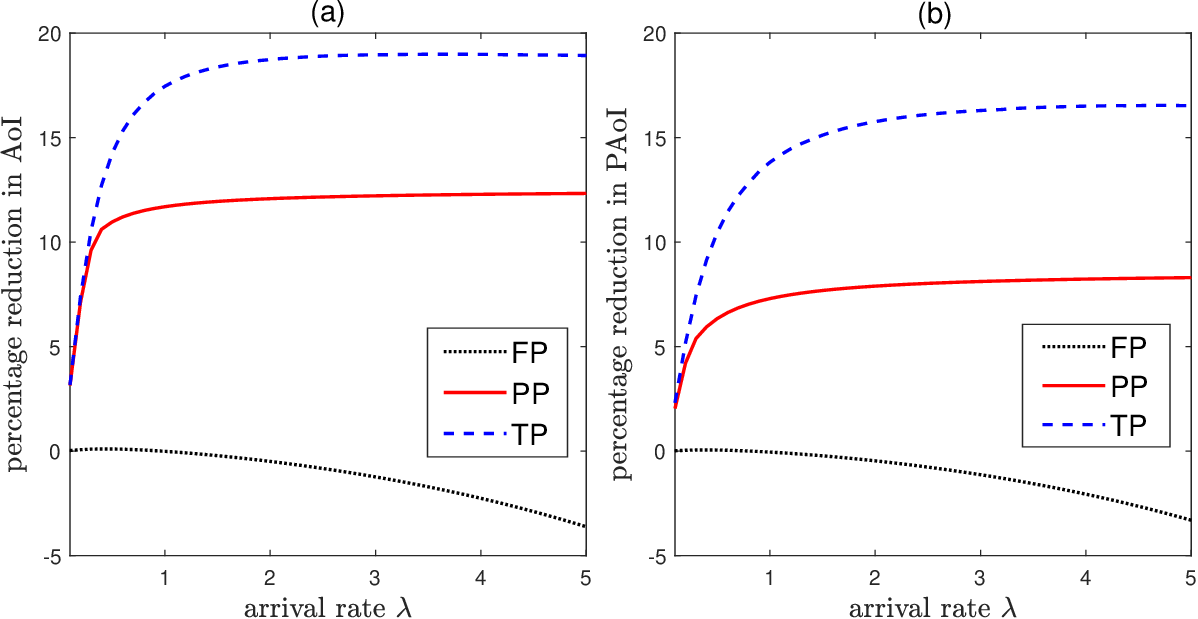}
    \caption{Percentage reduction in (a) average AoI (b) average PAoI, for the three policies FP, PP, and TP, the latter two policies tuned to perform optimally using Golden search. }
    \label{fig:PaperFig2}
\end{figure*}

 In this section, we study the average AoI and average PAoI under the PP and TP policies 
 using the preemption probability $\alpha$, and preemption threshold $\tau$, respectively, when the service time $S$ is log-normal distributed with pdf 
 \begin{align}
  f_S(t) & = \frac{1}{t \sigma \sqrt{2 \pi}} 
  {\rm e}^{-\frac{(\log t - \mu)^2}{2 \sigma^2}}, 
 \end{align}
 for $t > 0, \mu \in (-\infty,\infty), \sigma >0$.    
The service time mean and variance are written as,
\begin{align}
\mathbb{E}[S] & = {\rm e}^{\mu+\frac{\sigma^2}{2}}, \quad \text{Var}(S) = 
{\rm e}^{2\mu+\sigma^2} ( {\rm e}^{\sigma^2}-1 ).   
\end{align}
The service time parameters are fixed to $\mu = 0.75$ and $\sigma = 0.75$ as in \cite{moltafet_etal_isit25}.
Fig.~\ref{fig:PaperFig1}a  (resp. Fig.~\ref{fig:PaperFig1}b depicts the average AoI  
for the PP policy (resp. TP policy) as a function of the preemption probability $\alpha$ (resp. preemption threshold $\tau$), for four different values of the update arrival rate $\lambda$. Similarly, the average PAoI is depicted for both policies in Fig.~\ref{fig:PaperFig1}c and Fig.~\ref{fig:PaperFig1}d. 
We have the following observations:
\begin{itemize}
    \item For PP, the average AoI is a unimodal function 
    (see \cite{AppliedNumericalAnalysis}) of the preemption probability $\alpha$, i.e., there is exactly one point $\alpha^*$ in the interval $\alpha \in [0,1]$ for which the average AoI takes its minimum value, and when $\alpha < \alpha^*$ (resp. $\alpha > \alpha^*$), average AoI is a strictly decreasing function (resp. strictly increasing function) of $\alpha$. Similar observations can be made for the average PAoI under PP.
    \item  There are efficient search algorithms for finding the minimum of unimodal functions, such as the Golden search (see, for example, \cite{NumericalRecipes,AppliedNumericalAnalysis}), which we propose to use in this paper.
    \item  Also, for TP, average AoI is a unimodal function of the preemption threshold $\tau \in [0,\infty)$ which enables us to find the threshold $\tau^*$ giving rise to the lowest average AoI, again using the Golden search algorithm.
    \item For PP, the optimum preemption probability $\alpha^*$ decreases with increased $\lambda$. On the other hand, for TP, the optimum preemption threshold $\tau^*$ increases with increased $\lambda$.
    \item TP outperforms PP when the optimum threshold and probability parameters for both are employed. 
\end{itemize}
In the second numerical example, we study the same service time distribution and vary the update arrival rate $\lambda$ for each value of which we obtain the optimum preemption probability $\alpha^*$ for PP, and the optimum preemption threshold $\tau^*$ for TP, both using the Golden search algorithm. The percentage reduction in average AoI and average PAoI with the 
tuned PP and TP policies, along with the FP policy, with respect to the benchmark NP policy, are depicted in Fig.~\ref{fig:PaperFig2}. The results show that TP outperforms the two preemptive policies FP and PP, for all values of $\lambda$, whereas we have observed reductions up to $18.99\%$ and $16.54 \%$ with respect to the benchmark non-preemptive policy NP in terms of average AoI and average PAoI, respectively. We have also observed that the FP policy outperforms the NP policy in both metrics, except at very low arrival rates. 
Another observation is that the gap between the TP and PP policies is wider for the average PAoI metric compared to the average AoI. 

 \section{Conclusions}
 We study a single-source single-server status update system with generally distributed service times. For the case of log-normal distributed service times, the well-known practices of either not preempting at all or always preempting are shown to end up in poor performance, in terms of average AoI or average PAoI. We have also shown that probabilistic preemption and threshold-based preemption policies with the employment of optimally tuned preemption parameters result in significantly improved AoI/PAoI performance. 
 In particular, we have obtained up to $18.99 \%$ reduction in average AoI with the threshold-based policy in comparison to the non-preemptive policy, where the former outperformed all the other studied policies for all the scenarios investigated in this paper.
 \label{sec:conclusions}

\bibliographystyle{IEEEtran}
\bibliography{biblsinglesource}
\end{document}